\documentclass[twocolumn,showpacs,preprintnumbers,amsmath,amssymb]{revtex4}
\usepackage{dcolumn}
\usepackage{bm}
\usepackage{graphicx,xcolor}

\def\wo{\omega_{\text{osc}}}
\newcommand{\ket}[1]{\left| #1 \right\rangle } % for Dirac kets
\newcommand{\bra}[1]{\left\langle #1 \right| } % for Dirac kets
\def\wk{\omega_k}
\def\wj{\omega_j}
\def\wc{\omega_{\text{cut}}}

\begin{document}

\title{Field Fluctuations in a One-Dimensional Cavity with a Mobile Wall}

\author{Salvatore Butera and Roberto Passante}
\affiliation{Dipartimento di Fisica e Chimica, Universit\`{a} degli Studi di Palermo and CNISM, Via Archirafi 36, I-90123 Palermo, Italy}

\email{roberto.passante@unipa.it}


\begin{abstract}
We consider a scalar field in a one-dimensional cavity with a mobile wall. The wall is assumed bounded by a harmonic potential and its mechanical degrees of freedom are treated quantum mechanically. The possible motion of the wall makes the cavity length variable, and yields a wall-field interaction and an effective interaction among the modes of the cavity. We consider the ground state of the coupled system and calculate the average number of virtual excitations of the cavity modes induced by the wall-field interaction, as well as the average value of the field energy density. We compare our results with analogous quantities for a cavity with fixed walls, and show a correction to the Casimir potential energy between the cavity walls. We also find a change of the field energy density in the cavity, particularly relevant in the proximity of the mobile wall, yielding a correction to the Casimir-Polder interaction with a polarizable body placed inside the cavity. Similarities and differences of our results with the dynamical Casimir effect are also discussed.

\end{abstract}

\maketitle

We consider the zero-point field fluctuations inside a cavity with a mobile wall, treating quantum mechanically its mechanical degrees of freedom. We describe how the presence of a mobile wall affects vacuum fluctuations inside the cavity with respect to the usual case of a fixed wall. This is somehow related to the dynamical Casimir effect, that is the emission of real photons from the vacuum when a boundary condition of the field is set in motion with nonuniform acceleration \cite{Moore70,DMN11,Dodonov10}. A difference is that in our case, contrary to the dynamical Casimir effect, the motion of the wall is not prescribed, but it follows from the internal dynamics of the coupled wall-field system. The motion of the wall yields an effective wall-field interaction. We find that quantum fluctuations of the wall's position affect the field inside the cavity, as well as observable phenomena such as Casimir and Casimir-Polder forces.

Our model consists of a massless scalar field in a one-dimensional cavity with a fixed mirror at $x=0$ and a moving mirror with position $q=q(t)$. The moving mirror is described quantum mechanically, and it is also subjected to a harmonic potential $V(q)$. We are interested in studying how the motion of the cavity wall, in particular its ground-state fluctuations, affects the zero-point field fluctuations and related physical phenomena such as the Casimir force between the two walls and the Casimir-Polder force on a polarizable body placed inside the cavity. Nowadays cavities with walls with a very small mass can be experimentally obtained \cite{MG09,AKM13}, and the smaller the wall's mass, the larger the influence of its position fluctuations on the field. This kind of problems is related to the rapidly growing field of quantum optomechanics, which studies the coupling of field (optical) modes with mechanical degrees of freedom \cite{MG09,AKM13}. These studies also aim to build more sensitive force sensors, useful, for example, for gravitational wave detectors \cite{KV08}.
Even if this is not the main motivation of our work, we can compare our results due to the quantum fluctuations of the wall's position  with the thermal (Brownian) fluctuations considered in \cite{AD05,NKC04}. Thermal effects go to zero for vanishing temperatures, while our quantum effects are temperature independent and thus become more important for sufficiently low temperatures. A qualitative estimate, using the parameters in \cite{NKC04}, indicates that the quantum effects we obtain in this Letter are comparable with thermal effects for temperatures of a few kelvin, which appear at the reach of modern cryogenic laser interferometric detectors \cite{Uchiyama12}.

In our model, both the field and the moving mirror are described quantum mechanically. The massless scalar field $\phi(x,t)$ is described by the 1D Klein-Gordon equation
%\begin{equation}
%	\frac{\partial^2\phi(x,t)}{\partial x^2}-\frac 1{c^2}\frac{\partial^2\phi(x,t)}{\partial t^2}=0
%\label{eq:1}
%\end{equation}
for $0 \leq x \leq q(t)$, with the time-dependent boundary conditions $\phi (0,t)=\phi(q(t),t)=0$ for perfectly reflecting mirrors. The nonrelativistic equation of motion of the moving mirror of mass $M$ is
\begin{equation}
	M\ddot{q}(t)=- \frac {\partial V(q)}{\partial q} + \frac 12 \left. \left( \frac{\partial \phi(x,t)}{\partial x} \right)^2 \right|_{x=q(t)} \, .
\label{eq:2}
\end{equation}

Canonical quantization of our system has been obtained by Law in Refs. \cite{Law1,Law2}, in terms of an effective Hamiltonian yielding a coupling between the field modes. If $L_0$ is the equilibrium position of the mobile wall under the action of the harmonic potential $V(q)$, assuming small displacements from $L_0$, the Hamiltonian can be linearized and the effective wall-field interaction treated as a small perturbation. The linearized Hamiltonian is \cite{Law2}
\begin{equation}
	H=\frac {p^2}{2M}+\left( V(q)-\frac {\hbar c \pi}{24q} \right) +\hbar \sum_k {\omega_k a_k^\dag  a_k - x_mF_0} \, ,
\label{eq:3}
\end{equation}
where the first term is the kinetic energy of the mobile wall, the second its harmonic binding plus the Casimir interaction with the other wall, and the third term is the field Hamiltonian in terms of annihilation and creation operators for field modes relative to the mirror's equilibrium position $q=L_0$. Also, $x_m=q-L_0$ is the displacement of the wall from its equilibrium position, and
\begin{equation}
F_0=\frac \hbar{2L_0}\sum_{k,j}(-1)^{k+j}\sqrt{\omega_k \omega_j}\, {\cal N}\! \left[ \left( a_k + a_k^\dag \right) \left( a_j+
a_j^\dag \right)\right]
\label{eq:4}
\end{equation}
is the operator giving the radiation pressure on the wall, where ${\cal N}$ is the normal ordering operator. $k,j$ are integer numbers specifying the field modes in \eqref{eq:3} and \eqref{eq:4}, evaluated for the equilibrium position of the wall. The last term in \eqref{eq:3} is the mirror-field effective interaction energy.

Assuming that $V(q)$ is a harmonic potential with frequency $\wo$, and indicating with $b$ and $b^\dagger$ the bosonic operators for the mobile mirror, we can write the Hamiltonian \eqref{eq:3} in the form $H=H_0+H_I$, where
\begin{equation}
	 H_0=\hbar \wo b^\dag b +\hbar \sum_k \omega_k a_k^\dag  a_k
\label{eq:5}
\end{equation}
is the unperturbed Hamiltonian. The effective Hamiltonian describing the mobile mirror-field interaction is
\begin{equation}\begin{split}
H_{\text{int}}=&-\sum_{kj}C_{kj} \Big\{ \left( b + b^\dag \right) \\
&\times {\cal N} \! \left[ \left( a_k + a_k^\dag \right) \left(  a_j + a_j^\dag \right) \right] \Big\} \, ,
\label{eq:6}
\end{split}\end{equation}
where
\begin{equation}
C_{kj}=(-1)^{k+j}\left(\frac \hbar 2\right)^{3/2}\frac 1{L_0 \sqrt{M}}\sqrt{\frac{\omega_k \omega_j}\wo}
\label{eq:7}
\end{equation}
is the coupling constant. Equation \eqref{eq:6} includes an interaction between different field modes due to the motion of the wall.

We now consider the ground state of our system. The unperturbed ground state is $\ket{\left\{0_p\right\},0}$, where the first element refers to the field and the second one to the wall. This state is not an eigenstate of the total Hamiltonian because of the mirror-field interaction \eqref{eq:6}. The true ground state, at the lowest significant order in the wall-field interaction, is
\begin{equation}
	 \ket{g}=\ket{\left\{0_p\right\},0}+\sum_{k,j}{D_{kj}\ket{\left\{1_k,1_j\right\},1}} \, ,
\label{eq:8}
\end{equation}
where
\begin{equation}
D_{kj}=(-1)^{k+j}\frac 1{L_0} \sqrt{\frac {\wk \wj}{8\hbar M \wo}}\frac 1{\hbar\left(\wo+\wk+\wj\right)}.
\label{eq:9}
\end{equation}
It is possible to show that second-order terms in the coupling constant $D_{kj}$, as well as the normalization factor of the state \eqref{eq:8}, do not contribute, at the order considered, to the quantities we are interested in. For this reason we have not written them in \eqref{eq:8}. Equation \eqref{eq:8} shows that the interacting ground state contains terms with one excitation of the wall and two excitations in the field (that we shall call photons). This is analogous to the dynamical Casimir effect, where pairs of photons are emitted by an oscillating mirror  (see \cite{Dodonov10} and references therein). However, the two situations are conceptually different: in the dynamical Casimir effect the motion of the mirror is driven by an external action, while in our case it follows from the dynamics induced by Hamiltonian \eqref{eq:3}. Another striking difference is that in the dynamical Casimir effect real photons are emitted, while in our case the field excitations are virtual photons.

We now discuss two main features of the field in the cavity with the mobile wall: the photon spectrum and the field energy density, and discuss how they change, at the second order in the wall-field interaction,  with respect to the fixed wall case. In analogy with electromagnetic interactions, the energy density can be probed through the Casimir-Polder interaction with a polarizable body \cite{CPPP95}.

Let us first consider the photon spectrum due to the motion of the mirror. The average value of the photon number operator
$N_m = a_m^\dagger a_m$ of mode $m$ on the true ground state $\ket{g}$ given by \eqref{eq:8}, is
\begin{equation}
\bra{g}N_m\ket{g} = \sum_j \frac \hbar{2L_0^2 M}\frac {\omega_m \omega_j}\wo \frac 1{\left(\wo+\omega_m+\omega_j\right)^2} \, .
\label{eq:10}
\end{equation}

Also, the average value of the mirror excitation number operator $N_\text{osc} = b^\dagger b$ is

\begin{equation}
\bra{g}N_\text{osc}\ket{g} = \sum_{jk} \frac \hbar{4L_0^2 M}\frac {\omega_k \omega_j}\wo \frac 1{\left(\wo+\omega_k+\omega_j\right)^2} \, .
\label{eq:11}
\end{equation}

The field and mirror excitations in the {\it dressed} ground state \eqref{eq:8}, given by \eqref{eq:10} and \eqref{eq:11}, respectively, originate from the {\it bare} vacuum state $\ket{\left\{0_p\right\},0}$ as a consequence of the mirror-field interaction. Equation \eqref{eq:10} shows that $\bra{g}N_m\ket{g}$ is given by a sum of contributions proportional to $\omega_j \omega_m /[\wo (\wo +\omega_j+\omega_m)^2]$. As a function of $\omega_m$, it has a peak at $\wo + \omega_j$ and decreases with increasing mirror's oscillation frequency $\wo$ and mass $M$. This behavior has a clear physical meaning: a sort of interplay between the modes stimulates emissions of photons mainly with frequencies such that  $\omega_m \simeq \wo +\omega_j$, as if the wall oscillation and the photon of frequency $\omega_j$ foster excitation of a photon with frequency $\omega_m \simeq \wo +\omega_j$. Also, the higher the mirror oscillation frequency, the weaker its action in mediating the effective interaction between the modes and consequently the photon production in the cavity. An analogous consideration holds for the dependence on the mirror's mass.

Another remarkable property of \eqref{eq:11} is that the contribution of a single pair of field modes (i.e., each term in the sums in the equation) is maximum when the sum of the frequencies of the two virtual photons is equal to the mirror oscillation frequency, in analogy to the dynamical Casimir effect, where the total energy of the pair of emitted photons equals the cavity oscillation frequency times $\hbar$ \cite{MFMNR06}. In our case the (virtual) photons can have any frequency and they are not required to have a total energy equal to $\hbar \wo$, which is only the most probable value. This is related to the fact that the mirror's motion is not prescribed, but it follows the probabilistic laws of quantum mechanics.

The second-order energy shift due to the mirror-field interaction is
\begin{equation}
E_g^{(2)}=-\sum_{k j}\frac{\hbar^2}{4L_0^2 M} \frac {\omega_k \omega_j}\wo \frac 1{\left(\wo+\omega_k+\omega_j\right)}
%=& -\hbar \wo \bra{g}N_\text{osc}\ket{g} -\sum_k \hbar \omega_k \bra{g}N_k\ket{g} \, .
\label{eq:11a}
\end{equation}

This energy shift results in a change of the Casimir energy of the system, to be compared with the Casimir energy for fixed walls in \eqref{eq:3}. Its absolute value grows for decreasing mirror's mass and oscillation frequency, and can be relevant for small values of the mirror mass $M$.
This quantity needs a regularization, introducing a cutoff frequency $\wc$ in the frequency sums. The physical motivation for this cutoff is that a real mirror becomes transparent for frequencies larger than its plasma frequency \cite{AM76}, and thus its interaction with the field is strongly suppressed for such frequencies.

We can estimate the Casimir force change \eqref{eq:11a} with respect to the fixed walls case. The analogous Casimir force for fixed mirrors is of the order of $10^{-15} \, \text{N}$  for $L_0 = 1 \, \mu \text{m}$. Numerical evaluation of \eqref{eq:11a} shows that the correction to the force, which scales as $M^{-1}$, is around a few percent of this value when $M= 10^{-21} \text{kg}$, $\wo = 10^5 \, \text{s}^{-1}$ and a typical cutoff frequency $\wc = 10^{16}\, \text{s}^{-1}$. It can be further increased by reducing $\wo$. Even if this correction appears small, the possibility of measuring this tiny quantum effect relating mechanical and field degrees of freedom appears realistic in the near future; in fact, the actual precision of Casimir force measurements is around a few percent \cite{Lamoreaux05} and very small mass values can be achieved in quantum optomechanics experiments \cite{AKM13}.

A physical interpretation of the energy shift \eqref{eq:11a} can be obtained in terms of energy of emitted photons, mirror oscillation energy and mirror-field  interaction energy. In fact, we have
\begin{equation}
\bra{g} H_0 \ket{g} = \hbar \wo \bra{g}N_\text{osc}\ket{g} + \sum_k \hbar \omega_k \bra{g}N_k\ket{g} \, ,
\label{eq:12a}
\end{equation}
\begin{eqnarray}
\bra{g} H_{\text{int}} \ket{g} &=& -2\hbar \wo \bra{g}N_\text{osc}\ket{g} \nonumber \\
&-& 2\sum_k \hbar \omega_k \bra{g}N_k\ket{g} \, .
\label{eq:12b}
\end{eqnarray}
The sum of \eqref{eq:12a} and \eqref{eq:12b}, taking into account \eqref{eq:10} and \eqref{eq:11}, yields \eqref{eq:11a}, of course. This shows that the energy shift originates from a positive contribution of the energy stored in the field and in the oscillating mirror, and a negative contribution from the mirror-field interaction.

%\begin{figure}[h]\centering
%\includegraphics[width=8cm]
%{Fig1.eps}
%\caption{(Color online) Photon spectrum (arbitrary units) for the following values of the mirror's oscillation angular frequency: $5 \cdot 10^{4} \, %\text{s}^{-1}$ (black continuous line), $10^{5} \, \text{s}^{-1}$ (blue dashed line), $5 \cdot 10^{5} \, \text{s}^{-1}$ (green dot-dashed line). The equilibrium %cavity length and the cut-off angular frequency are $L_0 = 10 \mu \text{m}$ and $\wc = 10^{16}\, \text{s}^{-1}$, respectively.}
%\label{Fig:1}
%\end{figure}

\begin{figure}[h]\centering
\includegraphics[width=8cm]
{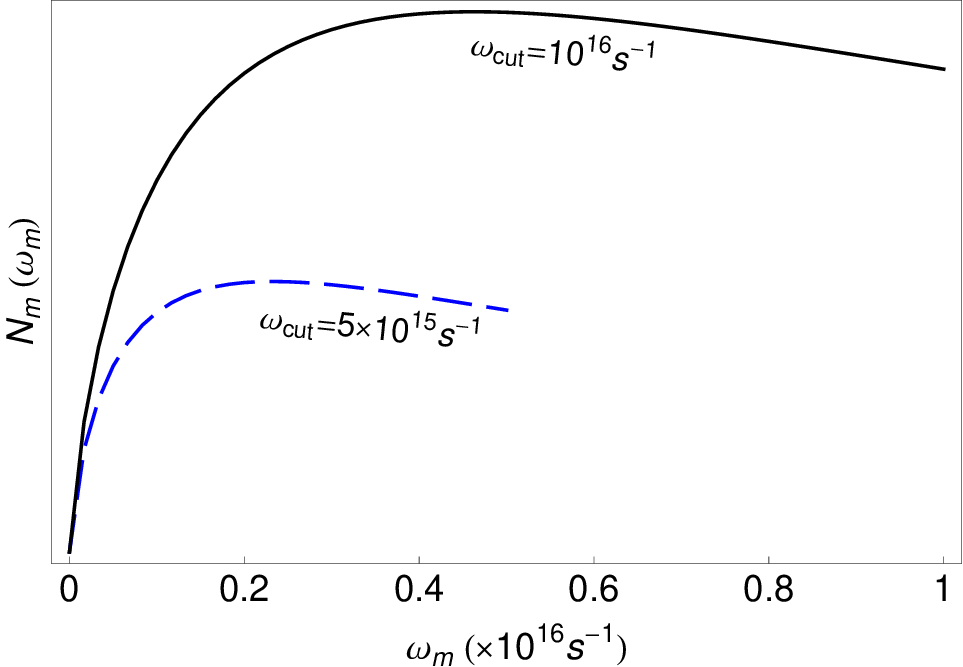}
\caption{(Color online) Photon spectrum (arb. units) for the following values of the cutoff angular frequency: $10^{16} \, \text{s}^{-1}$ (black solid line) and $5 \cdot 10^{15} \, \text{s}^{-1}$ (blue dashed line). The mirror's oscillation angular frequency is $\wo = 10^5 \, \text{s}^{-1}$ and the equilibrium cavity length is $L_0 = 10 \, \mu \text{m}$.}
\label{Fig:1}
\end{figure}

\begin{figure}[h]\centering
\includegraphics[width=8cm]
{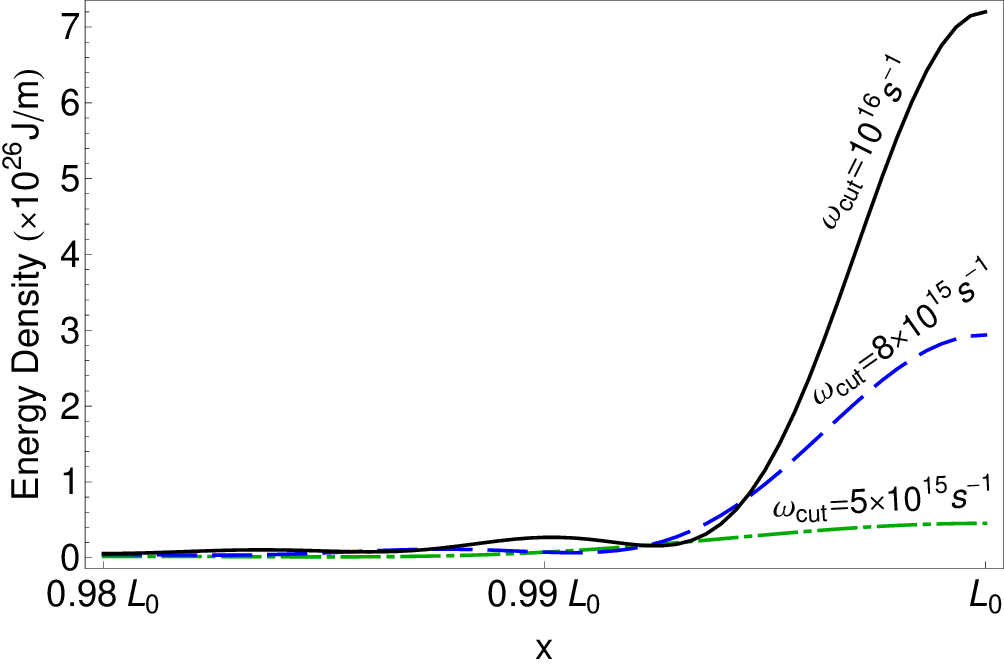}
\caption{(Color online) Change of the renormalized field energy density, compared to the static walls case, in the proximity of the moving mirror. The dependence on the cutoff frequency is evident, in particular close the moving mirror. Parameters are
$\wo = 10^5 \, \text{s}^{-1}$, $L_0 = 10 \, \mu \text{m}$ and $M=10^{-11}\, \text{kg}$.}
\label{Fig:2}
\end{figure}

Figure \ref{Fig:1} shows the photon number inside the cavity as a function of frequency for two different values of the cutoff frequency.
%oscillator's frequency $\wo = 10^5 \, s^{-1}$ and cavity length $L_0 = 10 \mu \text{m}$.
The values of the photon number depend on the mirror's mass $M$ as $1/M$. A typical mass of a commercial MEMS is $M=10^{-11}$ kg \cite{GZGLC13}, but much smaller masses in the range $10^{-15}-10^{-21}$ kg can be obtained nowadays in appropriate devices \cite{AKM13,SR05,MG09,NBG11}. The total number of photons in the cavity is obtained by integration of the curve in Fig. \ref{Fig:2}. It is proportional to $1/M$, and spans from $10^{-14}$ for the mass $M=10^{-11} \, \text{kg}$ of a commercial MEMS to $10^{-4}$ for masses of the order of $10^{-21} \, \text{kg}$, that are in reach of modern optomechanical technology \cite{AKM13}. Although these photon numbers may appear very small, they are similar to those involved in other observable quantum-electrodynamical effects such as the Lamb shift or atom-wall and atom-atom Casimir-Polder forces. Thus, the small average photon number does not imply at all that the position fluctuations of the wall have a negligible effect.

We now evaluate the field energy density inside the cavity, and investigate how it is affected by the motion of the mirror. The importance of considering this quantity is that, in analogy with electromagnetic interactions, it is related to observable quantities such as Casimir-Polder forces on polarizable bodies placed inside the cavity \cite{CPP95,CPPP95}.

The field energy density is given by
\begin{equation}
\mathcal{H}=\frac 12 \left[\frac 1{c^2} {\dot{\phi}}^2+\left( \frac {d {\phi}}{dx} \right)^2 \right] \, .
\label{eq:13}
\end{equation}
In the case of fixed mirrors, the renormalized ground-state energy density, i.e., after subtraction of the energy density  in the absence of the cavity walls, is
\begin{equation}
\langle \left\{0_p\right\} \mid \mathcal{H}^{\text{R}} \mid \left\{0_p\right\} \rangle =\langle  \mathcal{H}^{R}\rangle_0 =-\frac {\pi  c \hbar}{24 L_0^2}
\label{eq:14}
\end{equation}

In the case of a cavity with a mobile wall, we can calculate the expectation value of \eqref{eq:13} on the true ground state \eqref{eq:8}. After some algebra, subtracting the spatially homogeneous energy density present even in the absence of the cavity, we get the expression of the renormalized energy density on the ground state of the mobile wall-field interacting system
\begin{equation}
\langle g \mid \mathcal{H}^{\text{R}} \mid g \rangle =
\langle  \mathcal{H}^{R}\rangle_0 +\Delta \mathcal{H}
\label{eq:15}
\end{equation}
where $\langle  \mathcal{H}^{R}\rangle_0 $ is the same contribution \eqref{eq:14} for fixed walls, and
\begin{equation}
\begin{split}
\Delta \mathcal{H} &=\sum_j\sum_{kp} (-1)^{k+p} \frac{\hbar^2}{2L_0^3 M \wo} \\
&\times \frac {\omega_k \omega_j\omega_p}{(\wo+\omega_k+\omega_j)(\wo +\omega_p+\omega_j)}
\cos [(k-p)x]
\label{eq:16}
\end{split}
\end{equation}
is the change due to the motion of the wall, with respect to the fixed wall case. Even if a fixed wall is not realistic, the comparison can be meaningfully done with respect to a much more massive wall, by taking into account that our effect is proportional to $1/M$.

For the same reasons discussed after Eq. \eqref{eq:11a}, we introduce an upper cutoff frequency $\wc$ to regularize the ultraviolet divergence in the sum over $j$ of Eq. \eqref{eq:16}. We use a sharp cutoff to simulate the effect of a real material, which is equivalent to setting the number of field modes (in our 1D model all modes are equally spaced). The quantity \eqref{eq:16} can be evaluated numerically.

Figure \ref{Fig:2} shows the energy density in the proximity of the mobile wall, where the effect of its movement is more important. Parameters are the same as those of Fig. \ref{Fig:1}, with $M=10^{-11}\, \, \text{kg}$. The energy density is plotted for values of the cutoff frequency equal to typical plasma frequencies of a metal. The figure shows that including the motion of the wall affects the field energy density inside the cavity. This new effect is particularly relevant near the moving wall. This is consistent with the following physical picture: the virtual quanta generated by the wall remain confined in its proximity, similarly to the virtual photons around a ground-state atom, and are responsible for interatomic van der Waals or Casimir-Polder interactions \cite{CPP95}. Also, the figure clearly shows  that when the cutoff frequency is increased, the energy density has a strong increase in the very proximity of the wall. It diverges for $\wc \to \infty$; this is related to the known surface divergences of zero-point fluctuations at the interface between an ideal metallic mirror and the vacuum space \cite{MCPW06,BP12}. The presence of surface divergences of the energy density is a very important point, because they could act as a source for gravity \cite{EFM12,FR05}. It has been suggested that the problem of surface divergences at the interface could be solved by imperfect or fluctuating boundaries \cite{FS98}; although in our case the boundary indeed has quantum position fluctuations, the singular behavior of the energy density for $\wc \to \infty$ is still present because we have quantized the field in terms of operators relative to the wall's equilibrium position. We shall address this intriguing question in a successive paper. An estimate of the importance of our effect can also be obtained by comparing the energy density change near the mobile wall with the same quantity far from the mobile wall, for example in the middle of the cavity $x=L_0/2$. An explicit evaluation of \eqref{eq:16} at these two positions (parameters as in Fig.  \ref{Fig:2}, with $\wc = 10^{16} \, \text{s}^{-1}$) shows that the effect near the mobile wall is much larger (about 300 times) than at $x=L_0/2$.
The change of the energy density \eqref{eq:16} can be experimentally probed through the modification of the Casimir-Polder interaction on a polarizable body; our results clearly indicate that this effect can be important if the polarizable body is placed near the mobile wall and for very small mirror's masses.
This is a new effect arising from the effective interaction between the field and the wall's mechanical degrees of freedom.

To conclude, we have considered a scalar field in a one-dimensional cavity with one fixed and one mobile wall; the latter is bound to an equilibrium position by a harmonic potential and its mechanical degrees of freedom are treated quantum mechanically. The presence of the moving wall yields an effective interaction between the field modes. We have analyzed the ground state of the wall-field interacting system, which contains pairs of virtual photons, and discussed the photon spectrum and the field energy density. We have found modifications of the Casimir energy and of  the energy density near the mobile wall, with respect to the fixed wall case. The effect of these new phenomena on the Casimir interaction between the two walls, and on the Casimir-Polder interaction on a polarizable body, has been discussed in detail, as well as the possibility of observing them. Analogies and differences with the dynamical Casimir effect have been also discussed.

\begin{acknowledgments}
Financial support by the Julian Schwinger Foundation, by Ministero dell'Istruzione, dell'Universit\`{a} e della
Ricerca and by Comitato Regionale di
Ricerche Nucleari e di Struttura della Materia is
gratefully acknowledged. The  authors also acknowledge support from the ESF Research Networking Program CASIMIR.
\end{acknowledgments}

\end{document}